\documentstyle[12pt,iopconf]{article}

\begin{document} 

\bibliographystyle{unsrt}

\input epsf

\title{Representations of the discrete inhomogeneous Lorentz group and Dirac wave equation on the lattice}


\author{Miguel Lorente ${}^{1,2}$  and  Peter Kramer ${}^2$}

\affil{${}^1$ Departamento de F\'{\i}sica, Universidad de Oviedo, 33007 Oviedo, Spain}

\affil{${}^2$ Institute f\"ur theoretische Physik. Universit\"at T\"ubingen, Germany}


\beginabstract
We propose the fundamental and two dimensional representation of the Lorentz groups on a (3+1)-dimensional
hypercubic lattice, from which representations of higher dimensions can be constructed. For the unitary
representation of the discrete translation group we use the kernel of the Fourier transform. From the Dirac
representation of the Lorentz group (including reflections) we derive in a natural way the wave equation on the
lattice for spin 1/2 particles. Finally the induced representation of the discrete inhomogeneous Lorentz group is
constructed by standard methods and its connection with the continuous case is discussed.

\endabstract

\section{Introduction}

The hypothesis of a discrete space and time has attracted the attention of physicists for different
reasons [1] [3]:

i) as a mathematical tool in order to get rid of the infinities in quantum field theories with the help of a
cut-off in momentum space or, what is equivalent, a lattice for the space-time coordinates [7]. This model is similar
to solid state physics where quantum fields are defined on grid points of a periodic crystal.

ii) as a more realistic interpretation of lattice gauge theories, in which the space-time variables are
constrained to discrete values due to some  underlying structur resulting out of relations among
fundamental processes, as Penrose, Finkelstein, Weizsaecker have proposed. [3]

In this paper we address ourselves to an important problem of symmetries in lattice theories. In
particular we study the consequences of restricting the continuous space-time variables to a discrete
Minkowski space for the translations, rotations and Lorentz transformations. We discuss the possibility of
maintaining the representation theory for these groups on the lattice.

Our paper is based on the standard theory of induced representations of the Poincar\'e group restricted to
discrete variables [13]. In momentum space there are three ways to construct induced representations: Mackey,
Wigner and covariant states [10], based on the existence of a closed subgroup of a Lie group. In our case the
closed subgroup is the cubic group with respect to the Lorentz transformations on the lattice. We have the
advantage that the representations of the rotation group in two and three dimensions remain irreducible when
restricted to the cubic group. Therefore all the arguments for the discrete case can be taken unchanged from the
continuous case.

In a preliminary version [15] of this work we have stressed the connection of Klein-Gordon, Dirac and Proca
equation in discrete/continuous momentum and discrete space via Fourier transform.

In this paper we emphasize the representation theory of the discrete translation, Lorentz and Poincar\'e group in
such a way that the wave equaton for spin 1/2 particles on the lattice emerges in a natural way from the Direc
representation of the Lorentz group.

In section 2 we describe an algorithm to construct all integral transformations of the complete
Lorentz group based on the generators of some Coxeter group, and calculate the 2-dimensional
representations of this group that can be generalized to higher dimensional irreducible representations.

In section 3 we review two unitary irreducible representations of the discrete translation group and the
cyclic group and use these representations as the kernel of two Fourier transform on the lattice that have
become very helpful throughout the literature.

In section 4 we review the Dirac representations of the Lorentz group including space inversion and
construct the Dirac wave equation in momentum space as the projection operators that reduce the covariant
states of the representation to the irreducible components.

In section 5 we apply the Fourier transforms of section 3 to the Dirac equation in momentum space and
obtain a difference equation for the Dirac and Klein-Gordon fields on the lattice.

In section 6 we construct the induced representation of the Poincar\'e group on the lattice using the
Mackey-Wigner approach and discuss the irreducibility and orbit conditions of this representation.

\section{Fundamental and spin representation of the Lorentz group on the lattice}

An integral Lorentz transformation belongs to {\it GL(4,Z)} and leaves invariant the bilinear form
\begin{equation}
d^2={x}_{0}^{2}-{x}_{1}^{2}-{x}_{2}^{2}-{x}_{3}^{2}\,.
\end{equation}

According to Coxeter [1, page 47] all integral Lorentz transformations (including reflections) are obtained by
combining the operations of permuting the spatial coordinates ${x}_{1},{x}_{2},{x}_{3}$ and changing the
signs of any of the coordinates ${x}_{0},{x}_{1},{x}_{2},{x}_{3}$ together with the operation of adding the
quantity ${x}_{0}-{x}_{1}-{x}_{2}-{x}_{3}$ to each of the four coordinates of a point.

These operations can be described geometrically by the Weyl reflections on the planes perpendicular to the vectors
\[{\alpha }_{1}={e}_{1}-{e}_{2},\ {\alpha }_{2}={e}_{2}-{e}_{3},\ {\alpha
}_{3}={e}_{3},\ {\alpha }_{4}=-\left({{e}_{0}+{e}_{1}+{e}_{2}+{e}_{3}}\right)\,,\]
where $\left\{{{e}_{0},{e}_{1},{e}_{2},{e}_{3},}\right\}$ is an orthonormal basis.

In matrix form these reflections are

\[\begin{array}{ll}
{S}_{1}=\left({\begin{array}{cccc}1&0&0&0\\ 0&0&1&0\\ 0&1&0&0\\ 0&0&0&1\end{array}}\right),\
&{S}_{2}=\left({\begin{array}{cccc}1&0&0&0\\ 0&1&0&0\\ 0&0&0&1\\ 0&0&1&0\end{array}}\right),
\\
 {S}_{3}=\left({\begin{array}{cccc}1&0&0&0\\ 0&1&0&0\\ 0&0&1&0\\
0&0&0&-1\end{array}}\right),\ &{S}_{4}=\left({\begin{array}{cccc}2&1&1&1\\ -1&0&-1&-1\\
-1&-1&0&-1\\ -1&-1&-1&0\end{array}}\right).
\end{array}\]

These reflections generate a Coxeter group, the Dynkin diagram of which is the following



\hskip 25mm \epsfbox{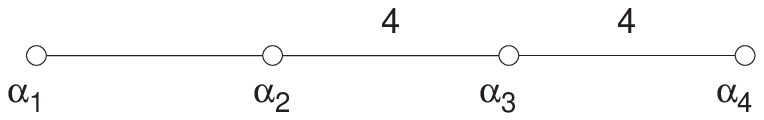}

\vskip 7pt

Fig.1: Dynkin diagram for the Coxeter group which generates all integral Lorentz transformations.

Kac [2] has proved that ${S}_{1},{S}_{2},{S}_{3},{S}_{4}$ generate all the integral Lorentz tranformations
that keep invariant the upper half of the light cone. Note that ${S}_{1},{S}_{2},{S}_{3}$ generate the full cubic or
octahedral group.

The generator ${S}_{1}$ can be used to factorize any integral Lorentz transformation that belongs to the complete
(orthocronus) Lorentz group. Let

\[
L\equiv \left({\begin{array}{cccc}\mit a&e&f&g\\ b&&&\\ c&&{}^*&\\
d&&&\end{array}}\right)
\]
be an integral matrix of determinant $\Delta =\pm 1$, satisfying
\begin{equation}
{L}^{t}gL=g\ ,\ g={diag}\left({ 1,-1,-1,-1}\right)
\end{equation}
and also $a\ge  1$. From
\[
{a}^{2}-{b}^{2}-{c}^{2}-{d}^{2}=1
\]
it follows that only one of $b,c,d$ can be zero. Suppose $a>1$. Then we apply
${S}_{1},{S}_{2},{S}_{3}$ to $L$ from the left until $b,c,d$ become non-positive integers. To the resulting
matrix we apply ${S}_{4}$. We get
\[L'\mit =\left({\begin{array}{cccc}a'&\mit e'&\mit f'&\mit g'\\
\mit b'&&&\\
\mit c'&&\mit {}^*&\\ d'&&&\end{array}}\right),\]
with $a'\mit =2a+b+c+d$. Obviously 
\[\left({a+b+c+d}\right)\left({a-b-c-d}\right)=1-2bc-2bd-2cd<0\,,\]
therefore $a+b+c+d<0$ \ or 
\[
2a+b+c+d=a'<a.
\]

By iteration of the same algorithm we get
\[
a>a'>a''>\ldots >{a}^{\left({k}\right)}\ge 1\,.
\]

The last inequality is a consequence of the fact that $L$ and ${S}_{4}$ belong to the complete Lorentz
group. Following this process we get an integral matrix with ${a}^{\left({k}\right)}=1$ which is a
combination of ${S}_{1},{S}_{2},{S}_{3}$, giving all the 48 elements of the cubic group on the
lattice.

Therefore a general integral Lorentz transformation of the complete Lorentz group $L$ can be
decomposed as

\begin{equation}
L={P}_{1}^{\eta }{P}_{2}^{\theta }{P}_{3}^{\iota}{S}_{4}\ldots {S}_{4}{P}_{1}^{\rm
\delta }{P}_{2}^{\varepsilon }{P}_{3}^{\zeta }{S}_{4}\left\{{{S}_{1}^{\alpha
}{S}_{2}^{\beta }{S}_{3}^{\gamma }}\right\}_{\rm all \ permutations}
\end{equation}
where ${P}_{1}={S}_{1}{S}_{2}{S}_{3}{S}_{2}{S}_{1},\ {P}_{2}={S}_{2}{S}_{3}{S}_{2},\
{P}_{3}={S}_{3}$ are matrices which change sign of $b,c,d$ and $\alpha \mit ,\beta \mit
,\gamma \mit ,\rm
\delta \mit ,\varepsilon \mit ,\zeta \mit ,\eta \mit ,\theta
\mit ,\iota \mit \ldots =0,1$

In the continuous case the boosts characterize the quotient of the Lorentz group with respect to the subgroup of
rotations. These continuous boosts take the vector $(1,0,0,0)$, stable under rotations, to any point of the unit
hyperboloid in the forward light cone. In the discrete case the vector $(1,0,0,0)$ is stable under the full cubic
group acting on the space coordinates. The coset representatives of the integral Lorentz group with respect to the
cubic subgroup are obtained by taking in equation (3) the quotient with respect to all integral elements of the cubic
group denoted by $\left\{{{S}_{1}^{\alpha }{S}_{2}^{\beta }{S}_{3}^{\gamma }}\right\}_{\rm all \ permutations.}$
These coset representatives allow us to construct an algorithm to obtain any vector of the unit hyperboloid in the
forward light cone from the vector $(1,0,0,0)$. Let $(a,b,c,d)$ be any vector with integral components satisfying
$a^2-b^2-c^2-d^2=1,(a\geq 1)$. First we apply to this vector the parity operators $P_1,P_2$ or $P_3$ defined in
equation (3), in such a way that the components $b, c, d$ become non-positive numbers. To the resulting vector we
apply the operator $S_4$ obtaining a new vector $(a',b',c',d')$, ${a'}^{2}-{b'}^{2}-{c'}^{2}-{d'}^{2}=1$, with
$a'=2a+b+c+d<a$. Following the same procedure we obtain new integral vectors satisfying $a>a'>a''>\ldots
>{a}^{\left({k}\right)}\ge 1$. In the last step we obtain the vector $(1,0,0,0)$ as required. Taking the product
of the generators used in this algorithm in inverse order we get the coset representative that takes the vector
$(1,0,0,0)$ to the vector $(a,b,c,d)$, and we call this coset representative the integral boost.

A particular case of integral Lorentz transformations are the integral Lorentz transformations without rotations.
These can be obtained with the help of Cayley parameters [3] making $n=p=q=0$ and $m,r,s,t$ integers. We can have
the following cases:

\begin{enumerate}
\item [i)] ${m}^{2}-{r}^{2}-{s}^{2}-{t}^{2}=1$

\begin{equation}
L=\left({\begin{array}{cccc}2{m}^{2}-1&2mr&2ms&2mt\\
2mr&2{r}^{2}+1&2rs&2rt\\
2ms&2rs&2{s}^{2}+1&2st\\
2mt&2rt&2st&2{t}^{2}+1\end{array}}\right),
\end{equation}

\item [ii)] ${m}^{2}-{r}^{2}-{s}^{2}-{t}^{2}=2$

\begin{equation}
L=\left({\begin{array}{cccc}{m}^{2}-1&mr&ms&mt\\
mr&{r}^{2}+1&rs&rt\\
ms&rs&{s}^{2}+1&st\\
mt&rt&st&{t}^{2}+1\end{array}}\right),
\end{equation}

\item [iii)] ${m}^{2}-{r}^{2}-{s}^{2}-{t}^{2}=-1$

\begin{equation}
L=\left({\begin{array}{cccc}{-2m}^{2}-1&2mr&2ms&2mt\\
2mr&{-2r}^{2}+1&-2rs&-2rt\\
2ms&-2rs&{-2s}^{2}+1&-2st\\
2mt&-2rt&-2st&{-2t}^{2}+1\end{array}}\right),
\end{equation}

\item [iv)] ${m}^{2}-{r}^{2}-{s}^{2}-{t}^{2}=-2$

\begin{equation}
L=\left({\begin{array}{cccc}{-m}^{2}-1&mr&ms&mt\\
mr&{-r}^{2}+1&-rs&-rt\\
ms&-rs&{-s}^{2}+1&-st\\
mt&-rt&-st&{-t}^{2}+1\end{array}}\right).
\end{equation}
\end{enumerate}

 The solutions of the diophantine equations in the four cases are obtained by the application of all the Coxeter
reflections as given in equation (3) to the vector $(1,0,0,0)$ in case i), to the vector $(2,1,1,0)$ in case ii), to
the vector  $(0,1,0,0)$ in case iii) and to the vector $(0,1,1,0)$ and $(1,1,1,1)$ in case iv). See Kac [2 page
70]. The four cases correspond to the integral Lorentz transformations given by Schild [1, page 42] restricted to the
pure Lorentz transformations without rotations.

Equations (4-7) can be considered also as particular cases of integral boost, that take the vector $(1,0,0,0)$ to
the vector defined by the first column of each of the four matrices. In the continuous case M{\o}ller [4] has given
a general boost that takes the vector $(1,0,0,0)$ to any vector on the unit hyperboloid. If we want to have a
general matrix that takes the vector $(1,0,0,0)$ to any integral vector $(M,R,S,T)$ on the unit hyperboloid this is
achieved by the square root of the first matrix of equation (4):

\begin{equation}
\sqrt {L}=\left({\begin{array}{cccc}M&R&S&T\\
R&1+{\frac{{R}^{2}}{1+M}}&{\frac{RS}{1+M}}&{\frac{RT}{1+M}}\\
S&{\frac{RS}{1+M}}&1+{\frac{{S}^{2}}{1+M}}&{\frac{ST}{1+M}}\\
T&{\frac{RT}{1+M}}&{\frac{ST}{1+M}}&1+{\frac{{T}^{2}}{1+M}}\end{array}}\right).
\end{equation}

This matrix is the restriction of the general Lorentz transformation given by M{\o}ller [4] to the values
\[M={\frac{{p}_{0}}{{m}_{0}c}},\ R={\frac{{p}_{1}}{{m}_{0}c}},\ S={\frac{{p}_{2}}{{m}_{0}c}},\
T={\frac{{p}_{3}}{{m}_{0}c}}\]
satisfying $M^2-R^2-S^2-T^2=1$.

In position space the space-time coordinates of the lattice ${x}_{\mu }$ are integer numbers. They transform under
integral Lorentz transformations into integral coordinates. The same is true for the increments $\Delta {x}_{\mu }$.

In momentum space the components of the four-momentum are not integer numbers but they can be constructed with the
help of integral coordinates, namely, 

\begin{equation}
{p}_{\mu }={m}_{0}c\ \left({{\frac{c\Delta t}{{\left({{\left({c\Delta t}\right)}^{2}-{\left({\Delta
\vec{x}}\right)}^{2}}\right)}^{1/2}}},\ {\frac{\Delta \vec{x}}{{\left({{\left({c\Delta \rm
t}\right)}^{2}-{\left({\Delta \vec{x}}\right)}^{2}}\right)}^{1/2}}}}\right).
\end{equation}

If $\Delta {x}_{\mu }$ transform under integral Lorentz transformations as a four-vector, the ${p}_{\mu }$
will transform also as a four-vector because the denominator in each component is Lorentz invariant.

In general there is no constraint between the values $\Delta {x}_{\mu }$. But if we impose the condition
${\left({c\Delta t}\right)}^{2}-{\left({\Delta \vec{x}}\right)}^{2}=1$ then  the four-momentum can be
considered an integral four vector multiplied by ${m}_{0}c$, ${m}_{0}c(M,R,S,T)$, satisfying
$(M^2-R^2-S^2-T^2)=1$ with $M=c\Delta t$, $R=\Delta x_1$, $S=\Delta x_2$, $T=\Delta x_3$.

Using the homomorphism between the groups {\it SO(3,1)} and {\it SL(2,C)} we obtain the representation of integral
Lorentz transformations as 2-dimensional complex matrices. From the knowledge of the Cayley parameters [5] for a
general element of the proper Lorentz group, we read off the matrix elements of the corresponding $\alpha \in \rm
SL\left({2,C}\right)$

\begin{equation}
\alpha ={\frac{1}{\sqrt {\Delta }}}\ \left({\begin{array}{cc}m+t+i\left({n-\lambda
}\right)&-p+r+i\left({q+s}\right)\\ p+r+i\left({q-s}\right)&m-t+i\left({n+\lambda }\right)\end{array}}\right),
\end{equation}
\begin{eqnarray*}
{\rm with}\quad \Delta &=&\det\ \alpha ={m}^{2}-{r}^{2}-{s}^{2}-{t}^{2}+{m}^{2}+{p}^{2}+{q}^{2}-{\lambda }^{2},\\
m\lambda &=& nt+ps+qr.
\end{eqnarray*}

For instance we can calculate the 2-dimensional representation of the Coxeter reflections $S_1$, multiplied by the
parity operator $P$ (in order to get an element of the proper Lorentz group) identifying its matrix elements with the
Lorentz matrix given in terms of Cayley parameters. Easy calculations give the unique solutions:

\begin{eqnarray}
\alpha \left({P{S}_{1}}\right) &=&\pm {\frac{1}{\sqrt {2}}}\left({\begin{array}{cc}0&-1-i\\ 1-i&0\end{array}}\right), \\
\alpha \left({P{S}_{2}}\right) &=&\pm {\frac{1}{\sqrt {2}}}\left({\begin{array}{cc}i&1\\-1&-i\end{array}}\right), \\
\alpha \left({P{S}_{3}}\right) &=&\pm \left({\begin{array}{cc}i&0\\0&-i\end{array}}\right), \\
\alpha \left({P{S}_{4}}\right) &=&\pm {\frac{1}{\sqrt {2}}}\left({\begin{array}{cc}0&1-i\\-1-i&2i\end{array}}\right).
\end{eqnarray}

From these matrices one can prove the expresions for the representation of the algebra for the Coxeter
reflections, namely,
\[\alpha \left({{S}_{i}{S}_{j}}\right)=\alpha \left({P{S}_{i}}\right)\alpha \left(
P{S}_{j}\right)\ \ \ ,\ \ \ i\ne j\ \ \ ,\ \ \ i,j=1,2,3\]
\[\alpha \left({P{S}_{i}}\right)\alpha \left({P{S}_{i}}\right)=-1\ \ \ ,\ \ \ i=1,2,3\]
\begin{equation}\alpha \left({P{S}_{i}}\right)\alpha \left({P{S}_{4}}\right)=\alpha \left({{S}_{\rm
i}{S}_{4}}\right)\ \ \ ,\ \ \ i=1,2,3
\end{equation}

The elements $\alpha \left({P{S}_{i}}\right), \,\ \ i=1,2,3$ generate the so called octahedral binary or double
group [14] excluding the parity. Together with $\alpha (PS_4)$ they generate part of a binary version of the
Coxeter group.

The integral Lorentz transformations without rotations as given in eqs. (4) to (7) have two dimensional
representations taking $n=p=q=\lambda=0$  in eq. (10) and the choices

\begin{enumerate}
\item[i)] $m^2-r^2-s^2-t^2=1$,
\item[ii)] $m^2-r^2-s^2-t^2=2$,
\item[iii)] $m^2-r^2-s^2-t^2=-1$,
\item[iv)] $m^2-r^2-s^2-t^2=-2$.
\end{enumerate}
in eq. (4-7). In order to complete the picture we have to add the 2-dimensional representation of the matrix $\sqrt
{L}$ given in eq. (8) which turns out to be

\begin{equation}
\alpha \left(\sqrt
{L}\right)\equiv {\kappa} ={\frac{1}{\sqrt {2\left({M+1}\right)}}}\left({\begin{array}{cc}M+1+T&R-iS\\
R+iS&M+1-T\end{array}}\right)={\kappa}^+.
\end{equation}

The $2\times2$ matrix representation of the discrete vector $(M,R,S,T)$, $M^2-R^2-S^2-T^2=1$ can be obtained from
the momentum in the rest system $(1,0,0,0)$ transformed by the matrix (16), namely,

\begin{equation}
{\kappa} \left({\begin{array}{cc}1&0\\0&1\end{array}}\right){\kappa}^{+}=\left({\begin{array}{cc}M+T&R-iS\\
R+iS&M-T\end{array}}\right).
\end{equation}

\section{Representations of the translation group on the lattice and Fourier transforms}

In the continuous case the unitary irreducible representations of the translation group are one dimensional and are
given by the function

\begin{equation}
{D}^{k}\left({x}\right)=\exp\ \left(i2\pi xk\right)\ \ \ ,\ \ \ x\in R,
\end{equation}
where $k$ is a continuous parameter that characterizes the representation.

If we restrict the translations to a discrete set of points $x=j\varepsilon$, $j\in Z$, the representation
becomes

\begin{equation}
{D}^{k}\left({j}\right)=\exp\ \left(i2\pi kj\varepsilon\right)
\end{equation}
where $k$ is still continuous. This representation satisfies orthogonality relations

\begin{equation}
\int_{-1/2\varepsilon }^{1/2\varepsilon
}{\overline{D}}^{k}\left({j}\right){D}^{k}\left({j'}\right)dk={\frac{1}{\varepsilon }}{\delta }_{jj'}
\end{equation}
and completeness relations

\begin{equation}
\sum\limits_{j\ =\ -\infty }^{\infty } {\overline{D}}^{k}\left({j}\right){D}^{k'}\left({
j}\right)={\frac{1}{\varepsilon }}\delta \left({k-k'}\right).
\end{equation}

The parameter of the discrete translation group $k$ is defined in the fundamental domain $-{\frac{1}{2\varepsilon }}\le
k\le {\frac{1}{2\varepsilon }} $.

From this representation we construct the {\it Type I Fourier transform on the lattice}
\begin{equation}
\hat{F}\left({k}\right)=\sum\limits_{j\ =\ -\infty }^{\infty } \exp\ \left({2\pi ikj\varepsilon
}\right){F}{\left({j}\right)}
\end{equation}
\begin{equation}
{F}{(j)}=\int_{-1/2\varepsilon }^{1/2\varepsilon }\exp\ \left({-2\pi ikj\varepsilon
}\right)\hat{F}\left({k}\right)dk
\end{equation}

If we generalize the discrete translation group to a (3+1)-dimensional cubic lattice
\[\Gamma :\left\{{{x}_{\mu }={j}_{\mu }\varepsilon ,\ \ \ \mu =0,1,2,3,\ \ \ {j}_{\mu }\in \rm
Z}\right\}\]
with scalar product
\begin{equation}
k\cdot x={k}^{\mu }{j}_{\mu }\varepsilon =\left({k\cdot j}\right)\varepsilon\ ,
\end{equation}
then the representations become
\begin{equation}
{D}^{k}\left({{j}_{\mu }}\right)=\exp2\pi i\left({k\cdot j}\right)\varepsilon\,.
\end{equation}

These representations satisfy orthogonal relations, completeness relations and Fourier transform analogous to the
1-dimensional case.

We introduce the reciprocal group of discrete translations on the reciprocal lattice ${\Gamma }^{R}$ by
\[{\Gamma }^{R}:\left\{{{b}^{R}={b}_{\mu }{\frac{1}{\varepsilon }}\ \ ,\ \ {b}_{\mu }\in {\rm Z}^{\rm 4}}\right\}.\]

The representations of the translation group in the $(3+1)$-cubic lattice are invariant under the reciprocal group
\begin{equation}
{D}^{k}\left({{j}_{\mu }}\right)={D}^{k+{b}^{R}}\left({{j}_{\mu }}\right)\ ,
\end{equation}
therefore, the parameters of the representations $k$ are restricted to the fundamental domain
\[-{\frac{1}{2\varepsilon }}\ <\ {k}_{\mu }\ \le \ {\frac{1}{2\varepsilon }}\,.\]

The same property applies to the Fourier transform in (3+1)-dimensions
\begin{equation}
\hat{F}\left({k}\right)\ =\ \hat{F}\ \left.{\left({k
+{b}^{R}}\right)}\right.\,,\,{b}^{R}\ \in {\Gamma }^{R}.
\end{equation}

The irreducible representations of the reciprocal group can be written
\begin{equation}
{D}^{\xi }\left({{b}^{R}}\right)\ =\ \exp\ \left({2\ \pi \ i\ \xi\ \cdot \
{b}^{R}}\right),\ \,-\in /2 \ \le {\xi }_{\mu }\le \in /2 
\end{equation}
satisfying
\begin{equation}
\int_{-\varepsilon/2}^{\varepsilon/2}{\overline{D}}^{\xi}\left({{b}^{R}}\right){D}^{\xi
}\left({{b}^{R'}}\right)\ d^4\xi \ =\ \varepsilon^4 {\delta }_{b^R ,\ b^{R'}}
\end{equation}
\begin{equation}\sum\limits_{{b}^{R}\in {\Gamma }^{R}}^{\infty } {\overline{
D}}^{\xi }\left({{b}^{R}}\right)\ {D}^{\xi '}\left({{b}^{R}}\right)\ =\
\varepsilon^4
\delta \ \left({\xi -\xi '}\right)
\end{equation}

Now we combine the irreducible representions of the discrete translations group in (3+1)-dimensions and the integral
Lorentz transformations.

Given a periodic function on the $k$-space
\begin{equation}
\hat{F}\left({k}\right)=\hat{F}\left({k+{b}^{R}}\right),
\end{equation}
it can be written in terms of discrete waves
\begin{equation}
\hat{F}\left({k}\right)=\sum\limits_{{j}_{\mu } \ \in  \ \Gamma }^{} {\rm \exp}\left(2\pi
i\left({{k}\cdot{j}}\right)\varepsilon\right)  \ F\left({{j}_{\mu }}\right).
\end{equation}

From the action of the group we have
\[{U}_{\Lambda }\hat{F}\left({k}\right)=\hat{F}\left({{\Lambda }^{-1}k}\right).\]

Then
\[{U}_{\Lambda }\hat{F}\left({k+{b}^{R}}\right)=\hat{F}\left({{\Lambda
}^{-1}k+{\Lambda }^{-1}{b}^{R}}\right)\]

But from the properties of the scalar product and using (32) we have
\[\exp\left({2\pi i\left({{\Lambda }^{-1}{b}^{R}\cdot j}\right)\varepsilon }\right)=\exp2\pi 
i\left({{b}^{R}\cdot {\left({{\Lambda }^{\rm -1}}\right)}^{T}j}\right)\varepsilon  =\exp2\pi 
i\left( b_{\mu }^{R} \left( \Lambda^{-1}\right)_{\nu }^{T\mu }{j}^{\nu }\right)=1\]
for any integral Lorentz transformation. Finally, 
\begin{equation}
{U}_{\Lambda }\hat{F}\left({k+{b}^{R}}\right)=\hat{F}\left({{\Lambda }^{-1}k}\right)={U}_{\Lambda
}\hat{F}\left({k}\right).
\end{equation}

Therefore, the periodicity of some function with respect to the reciprocal group is conserved under  integral
Lorentz transformations.

Now we consider the one dimensional unitary irreducible representation of the cyclic group in one dimension of order
$N$:
\begin{equation}
{D}^{m}\left({j}\right)=\exp{\frac{2\pi \rm i}{N}}mj\ \ \ ,\ \ \ m,j=0,1,\cdots \rm ,N-1,
\end{equation}
where $j$ represents the space variable and $m$ the label of the representation.

This representation satisfies periodic boundary conditions with respect to the $j$ variable

\[{D}^{m}\left({j+N}\right)={D}^{m}\left({j}\right),\]
and also with respect to the $m$ label

\[{D}^{m+N}\left({j}\right)={D}^{m}\left({j}\right),\]
and satisfies orthogonality relations
\begin{equation}
{\frac{1}{N}}\sum\limits_{m\ =\ 0}^{N-1} {D}^{m}\left({j}\right){D}^{m}\left({j'}\right) ={\delta }_{mm'}
\end{equation}
and completeness relations
\begin{equation}
{\frac{1}{N\varepsilon }}\sum\limits_{ j\ =\ 0}^{ N-1} { D}^{ m}\left({ j}\right){ D}^{
m'}\left({ j}\right) ={\frac{1}{\varepsilon }}{\delta }_{jj'}
\end{equation}

From this representation we construct a {\it Type II Fourier transform on the lattice}
\begin{equation}
\hat{F}\left({m}\right)={\frac{1}{\sqrt {N}}}\sum\limits_{ j\ =\ 0}^{ N-1}  \exp\left({i{\frac{2\pi }{
N}}mj}\right)F\left({j}\right)
\end{equation}
for any periodic function of discrete variable $F(j+N)=F(j)$. This is called in the literature the Finite Fourier
transform.

The representation of the cyclic group in the one-dimensional lattice and the corresponding Fourier transform can be
generalized to the (3+1)-dimensional cubic lattice,
\begin{equation}
{D}^{m}\left({{j}_{\mu }}\right)=expi{\frac{2\pi }{ N}}\left({m \cdot j}\right),
\end{equation}
with $\left({m \cdot j}\right)={{m}^{\mu }j}_{\mu }$  the bilinear form invariant under integral Lorentz transformations.

Because of the periodic boundary conditions the label of the representation of the cyclic group on the lattice is
constrained to the fundamental domain, namely,
\[{D}^{m}\left({{j}_{\mu }}\right)={D}^{m}\left({{\xi }_{\mu }+{n}_{\mu }N}\right)={D}^{m}\left({{\xi }_{\mu
}}\right),\]
where $0\le {\xi }_{\mu }\le  N\ \ \ ,\ \ \ {\xi }_{\mu }\in  Z\ \ \ ,\ \ \ {n}_{\mu }\in  Z$

Symbolically
\begin{equation}
{D}^{m}\left({j}\right)={D}^{m}\left({\xi  +{j}^{c}}\right)={D}^{m}\left({\xi }\right),
\end{equation}
where $\xi$ belongs to the fundamental domain and $j^c$ is any vector in the cubic lattice whose components are
multiples of $N$, ${j}^{c}=\left\{{{n}_{\mu }N}\right\}$. The boundary conditions (39) are invariant under integral
Lorentz transformations. From the definition of the group action
\[{U}_{\Lambda }f\left({x}\right)=f\left({{\Lambda }^{-1}x}\right),\]
we have for a periodic function, periodic with respect to the space variable,
\[f\left({\xi  +{j}^{c}}\right)=f\left({\xi }\right),\]
\begin{equation}
\left({{U}_{\Lambda }f}\right)\left({\xi  +{j}^{c}}\right)=f\left({{\Lambda }^{-1}\xi  +{\Lambda
}^{-1}{j}^{c}}\right)=f\left({{\Lambda }^{-1}\xi }\right)=\left({{U}_{\Lambda }f}\right)\left({\xi }\right),
\end{equation}
where we have used the property
\[\exp\left(i{\frac{2\pi }{ N}}\sum\limits_{\rho  ,\mu }^{} {\Lambda }_{\mu }^{\rho }{ n}_{\rho } N{m}_{\mu
}\right)=1,\]
due to the integral character of the Lorentz transformations.

\section{Dirac representation of the Lorentz group and covariant states}

Let $ L\left({\alpha }\right)$ be an element of the proper Lorentz group corresponding to the
element $\alpha \in SL\left({2,C}\right)$ and $I_s$ the parity operator. One writes the components of
four-momentum as a $2\times 2$ matrix

\begin{equation}
\tilde{p}\equiv {p}^{\mu }{\sigma }_{\mu }={p}^{0}{\sigma
}_{0}+{p}^{i}{\sigma }_{i}\ ,
\end{equation}
where ${\sigma}_0=1$ and ${\sigma}_i$ are the Pauli matrices.

The transformations of $\tilde{p}$ under parity and $SL(2,C)$ are

\[{I}_{s}:\tilde{p}\rightarrow {\tilde{p}}^{s}={p}^{0}{\sigma }_{0}-{p}^{i}{\sigma
}_{i}=\left({\det\ \tilde{p}}\right){\left({\tilde{p}}\right)}^{-1}\ ,\]

\begin{equation}
\alpha :\tilde{p}\rightarrow \alpha \tilde{p}{\alpha }^{+}\ ,
\end{equation}

\[{\tilde{p}}^{s}\rightarrow {\left({{\alpha
}^{+}}\right)}^{-1}{\tilde{p}}^{s}{\alpha }^{-1}.\]

It follows
\begin{equation}
I_sL(\alpha)I_s^{-1} = L\left((\alpha^+)^{-1}\right).
\end{equation}

Therefore the matrix ${\left({{\alpha}^{+}}\right)}^{-1}$ gives another 2-dimensional
representation of the Lorentz group non-equivalent to $\alpha \in SL\left({2,C}\right)$.
In order to enlarge the proper Lorentz group with space reflection we take both representations
$\alpha$ and ${\left({{\alpha }^{+}}\right)}^{-1}$.

Let $\pi \equiv \left\{{\mit I,{I}_{s}}\right\}$ the space reflection group and $\alpha
\in SL\left({2,C}\right)$, then the semidirect product 

$$SL\left({2,C}\right)\otimes \pi$$
with the multiplication law

\begin{eqnarray}
\left({\alpha ,\pi }\right)\ \left({\alpha ',\pi '}\right)&=&\left({\alpha
\alpha ',\pi \pi '}\right)\ if\ \pi =I, \\
\left({\alpha ,\pi }\right)\ \left({\alpha ',\pi '}\right)&=&\left({\alpha
{({{\alpha '}^{+}})}^{-1},\pi \pi '}\right)\ if\ \pi ={I}_{s},
\end{eqnarray}
form a group.

This group has a 4-dimensional representation, particular elements of which are

\begin{equation}
\overline{D}\left({\alpha ,I}\right)=\left({\begin{array}{cc}\alpha
&0\\ 0&{\left({{\alpha }^{+}}\right)}^{-1}\end{array}}\right),\
\overline{D}\left({e,{I}_{s}}\right)=\left({\begin{array}{cc}0&{\sigma }_{0}\\ {\sigma
}_{0}&0\end{array}}\right)
\end{equation}
that satisfy

\begin{equation}
\overline{D}\left({e,{I}_{s}}\right)\overline{D}\left({\alpha
,I}\right)\overline{D}\left({e,{I}_{s}^{-1}}\right)=\
\overline{D}\left({{\left({{\alpha }^{+}}\right)}^{-1},I}\right).
\end{equation}

In this representation we could now construct $\overline{D}(S_i)$ for the generators of the Coxeter group. Using formulas
(11)-(14) we get the 4-dimensional matrices:
\begin{eqnarray*}
\overline{D}\left({{S}_{i}}\right)&=&\left({\begin{array}{cc}0
&\alpha (PS_i)\\ \alpha (PS_i)&0\end{array}} \right) \ \ \ i=1,2,3 \\ 
\overline{D}\left({{S}_{4}}\right)&=&\left({\begin{array}{cc}0
&\left[\alpha^+ (PS_4)\right]^{-1}\\ \alpha (PS_4)&0\end{array}} \right)
\end{eqnarray*}
\[\overline{D}\left({{S}_{i}}\right)\ \overline{D}\left({{S}_{i}}\right)=-1 \qquad \qquad i=1,2,3,4\]
\[\left[\overline{D}\left({{S}_{1}}\right)\ \overline{D}\left({{S}_{2}}\right)\right]^3=\pm 1\ ,\
\left[\overline{D}\left({{S}_{2}}\right)\ \overline{D}\left({{S}_{3}}\right)\right]^4=\pm 1\ ,\
\left[\overline{D}\left({{S}_{3}}\right)\ \overline{D}\left({{S}_{4}}\right)\right]^4=\pm 1\]

With respect to this representation, a 4-component spinor $\overline{\psi }\left({p}\right)$ in momentum space
transforms as follows:
\begin{eqnarray}
U\left({\alpha ,I}\right)\overline{\psi
}\left({p}\right)&=&\overline{D}\left({\alpha ,I}\right)\overline{\psi }\left({{L}^{-1}\left({\alpha
}\right)p}\right), \\
U\left({e,{I}_{s}}\right)\overline{\psi
}\left({p}\right)&=&\overline{D}\left({e,{I}_{s}}\right)\overline{\psi }\left({{I}_{s}p}\right).
\end{eqnarray}

Using a similarity transformation we obtain an equivalent representation

\[D\left({\alpha ,\pi }\right)=M\overline{D}\left({\alpha ,\pi }\right){M}^{-1}\]
with
\[M={\frac{1}{\sqrt {2}}}\left({\begin{array}{cc}{\sigma }_{0}&{\sigma }_{0}\\
{-\sigma }_{0}&{\sigma }_{0}\end{array}}\right).\]

In this representation

\begin{eqnarray}
D\left({\alpha ,I}\right)&=&{\frac{1}{2}}\left({\begin{array}{cc}{\alpha +\left({{\alpha
}^{+}}\right)}^{-1},&{\mit -\alpha +\left({{\alpha }^{+}}\right)}^{-1}\\ {\mit -\alpha
+\left({{\alpha }^{+}}\right)}^{-1}\mit ,&{\alpha +\left({{\alpha }^{+}}\right)}^{
-1}\end{array}}\right), \\
D\left({e,{I}_{s}}\right)&=&\left({\begin{array}{cc}{\sigma }_{0}&0\\
0&{\mit -\sigma }_{0}\end{array}}\right).
\end{eqnarray}

For this representation we can derive from eq. (46) the unitarity relation

\[{D}^{+}\left({\alpha  ,\pi }\right)D\left({e,{I}_{s}}\right)D\left({\alpha  ,\pi
}\right)=D\left({e,{I}_{s}}\right).\]

The new four-spinor

\[\psi \left({p}\right)=M\overline{\psi }\left({p}\right)\]

transforms as

\begin{eqnarray}
U\left({\alpha ,I}\right)\psi \left({p}\right)&=&D\left({\alpha ,I}\right)\psi
\left({{L}^{-1}\left({\alpha }\right)p}\right), \\
U\left({e,{I}_{s}}\right)\psi \left({p}\right)&=&D\left({e,{I}_{s}}\right)\psi
\left({{I}_{s},p}\right),
\end{eqnarray}
and has an invariant scalar product due to the unitary relation given above. We call this representation the Dirac
representation.

The Dirac wave equation can be considered as a consequence of the relativistic invariance and irreducibility[10]:
Under the restriction to $SU(2)$, the first and second pair of components of the four-spinor transform according to
spin 1/2. Irreducibility requires that one of these pairs should be eliminated. In the present discrete case we must
replace the continuous group $SU(2)$ by the binary octahedral group [14]. Fortunately the restriction of $SU(2)$ to
this discrete subgroup is irreducible [11]. This allows us to follow the steps of the continuous analysis [10]. In
the rest system we want a projection operator that selects one irreducible representation of $SU(2)$ out of the
Dirac representation. This is achieved in the rest system by the projection operator:

\begin{equation}
Q={\frac{1}{2}}\left({I+\beta }\right),\ \ \ \beta \equiv \left({\begin{array}{cc}{\sigma
}_{0}&0\\ 0&-{\sigma }_{0}\end{array}}\right).
\end{equation}

In order to get the projection operator in an arbitrary system we apply $D(\kappa, I)$ given by eqs.(50) and (16). In the last
equation if we identify $M={\frac{{P}_{0}}{{m}_{0}c}},\ R={\frac{{P}_{1}}{{m}_{0}c}},\ S={\frac{{P}_{2}}{{m}_{0}c}},\
T={\frac{{P}_{3}}{{m}_{0}c}}\ ,$ then,

\begin{equation} Q\rightarrow
Q\left({\kappa}\right)={D}^{-1}\left({\kappa},I\right)QD\left({\kappa},I\right)={\frac{1}{2}}\left({I+W\left({\kappa}\right)}\right),
\end{equation} where 
\begin{equation}
W\left({\kappa}\right)={\frac{1}{2}}\left({\begin{array}{cc}{\kappa}^{+}\kappa+{\left({{\kappa}^{+}\kappa}\right)}^{-1},&-{\kappa}^{+}\kappa+{\left({{\kappa}^{+}\kappa}\right)}^{-1}\\
{\kappa}^{+}\kappa-{\left({{\kappa}^{+}\kappa}\right)}^{-1},&-{\kappa}^{+}\kappa-{\left({{\kappa}^{+}\kappa}\right)}^{-1}\end{array}}\right)
\end{equation}

Using the identities

 \parbox{11cm}{ 
\begin{eqnarray*}
& &{\left({{\kappa}^{+}\kappa}\right)}^{-1}={\frac{1}{{m}_{0}c}}{\sigma }^{\mu }{p}_{\mu }\ , \\
& &{\kappa}^{+}\kappa={\frac{1}{{m}_{0}c}}{\sigma }^{0}{p}_{0}-{\sigma}^{j}{p}_{j},
\end{eqnarray*}}\hfill
\parbox{1cm}{\begin{eqnarray}\end{eqnarray}}

\noindent we find
\begin{equation}
W\left({\kappa}\right)={\frac{1}{{m}_{0}c}}{\gamma }^{\mu}{p}_{\mu }\ ,
\end{equation}
where ${\gamma }^{\mu}$ are Dirac matrices with the realization

\[{\gamma }^{0}=\left({\begin{array}{cc}{\sigma }^{0}&0\\
0&-{\sigma }^{0}\end{array}}\right)\ \ \ ,\ \ \ {\gamma }^{j}=\left({\begin{array}{cc}0&{\sigma }^{j}\\ -{\sigma }^{j}&0\end{array}}\right)\]

\[{\gamma }^{0}={\gamma }_{0}\ \ \ ,\ \ \ {\gamma }^{j}={-\gamma }_{j}.\]

Collecting these result we obtain the Dirac equation in momentum space

\[ Q\left(\kappa\right)\psi \left( p\right)=\frac{1}{2}\left(I+W\left(\kappa\right)\right)\psi
\left( p\right)=\psi \left( p\right)\]
or

\begin{equation}
\left({{\gamma }^{\mu }{p}_{\mu }-{m}_{0}cI}\right)\psi
\left({p}\right)=0.
\end{equation}

(An equivalent method can be used applying to the projection operator the Foldy-Wouthuysen
transformation [9])

We apply the operator ${\gamma }_{\mu }{p}^{\mu }+{m}_{0}c$ from the left to (59) and obtain the mass-shell condition

\begin{equation}
\left({{p}^{\mu }{p}_{\mu }-{m}_{0}^{2}{c}^{2}}\right)\psi
\left({p}\right)=0.
\end{equation}

The Dirac equation is invariant under the group with elements $\left({\alpha ,\pi }\right)$ defined before.
In other words, if $\psi \left({p}\right)$ is a solution of the Dirac equation, so is $
U\left({\alpha ,\pi }\right)\psi \left({p}\right)$. Put $\pi=I$. Then

$$U\left({\alpha ,I}\right)\psi \left({p}\right)=D\left({\alpha ,I}\right)\psi
\left({{L}^{-1}p}\right),$$

$${D}^{-1}\left({\alpha }\right)Q\left({\kappa}\right)D\left({\alpha
}\right)=Q\left({\kappa \alpha }\right),$$

$$Q\left({\kappa \alpha }\right)\psi \left({{L}^{-1}p}\right)=\psi \left({{L}^{-1}p}\right).$$

Therefore
\begin{eqnarray*}
Q\left({\kappa}\right)U\left({\alpha ,I}\right)\psi
\left({p}\right)&=&Q\left({\kappa}\right)D\left({\alpha ,I}\right)\psi \left({{L}^{-1}\left({\alpha
}\right)p}\right)= 
D\left({\alpha ,I}\right)Q\left({\kappa \alpha }\right)\psi \left({{L}^{-1}\left({\alpha
}\right)p}\right)=\\
&=& D\left({\alpha ,I}\right)\psi \left({{L}^{-1}\left({\alpha
}\right)p}\right)=U\left({\alpha ,I}\right)\psi \left({p}\right)
\end{eqnarray*}
as required. For the space reflection
$$Q\left({{I}_{s}\kappa}\right)\psi \left({{I}_{s}p}\right)=\psi \left({{I}_{s}p}\right),$$
$${D}^{-1}\left({e,{I}_{s}}\right)
Q\left({\kappa}\right)D\left({e,{I}_{s}}\right)=Q\left({{I}_{s}p}\right),$$
$$U\left({e,{I}_{s}}\right)\psi \left({p}\right)=D\left({e,{I}_{s}}\right)\psi
\left({{I}_{s}p}\right),$$
we get
\begin{eqnarray*}
Q\left({\kappa}\right)U\left({e,{I}_{s}}\right)\psi
\left({p}\right)&=&Q\left({\kappa}\right)D\left({e,{I}_{s}}\right)\psi
\left({{I}_{s}p}\right)=D\left({e,{I}_{s}}\right)Q\left({{I}_{s}\kappa}\right)\psi
\left({{I}_{s}p}\right)= \\
&=& D\left({e,{I}_{s}}\right)\psi
\left({{I}_{s}p}\right)=U\left({e,{I}_{s}}\right)\psi \left({p}\right)
\end{eqnarray*}
as required. Notice that, due to the relation between $SU(2)$ and the binary octahedral group mentioned above, all
properties of Dirac representation in continuous momentum space carry over to the discrete momentum space without
modification.

\section{Dirac and Klein-Gordon equation on the lattice}

From the Dirac equation in momentum space (59) we can construct the wave equation in position space with the help of the
Fourier transform we have introduced in section 3. We define the following difference operators
\begin{equation}
\Delta  f\left({j}\right)=f\left({j+1}\right)-f\left({j}\right),\ \ \tilde{\Delta
}f\left({j}\right)={\frac{1}{2}}\left\{{f\left({j+1}\right)+f\left({j}\right)}\right\},
\end{equation}
\begin{equation}
\nabla  f\left({j}\right)=f\left({j}\right)-f\left({j-1}\right),\ \ \tilde{\nabla
}f\left({j}\right)={\frac{1}{2}}\left\{{f\left({j}\right)+f\left({j-1}\right)}\right\},
\end{equation}
and the partial difference operators with respect to a function of several discrete variables
\[{\Delta }_{\nu }f\left({{j}_{\mu }}\right)=f\left({{j}_{\mu }+{\delta }_{\mu \nu }}\right)-f\left({{j}_{\mu }}\right),\]
\begin{equation}
{\tilde{\Delta }}_{\nu }f\left({{j}_{\mu }}\right)={\frac{1}{2}}\left\{{f\left({{j}_{\mu }+{\delta }_{\mu \nu
}}\right)+f\left({{j}_{\mu }}\right)}\right\},
\end{equation}
and similarly ${\nabla }_{\nu }f\left({{j}_{\mu }}\right)$ and ${\tilde{\nabla }}_{\nu }f\left({{j}_{\mu }}\right)$.

From these operators we construct
\begin{equation}
{\delta }_{\mu }^{+}\equiv {\frac{ 1}{\varepsilon }}{\Delta }_{\mu }\prod\limits_{\nu \ne \mu }^{} {\tilde{\Delta }}_{\nu }\
 \ ,\ \ {\delta }_{\mu }^{-}\equiv {\frac{ 1}{\varepsilon }}{\nabla }_{\mu }\prod\limits_{\nu \ne \mu }^{} {\tilde{\nabla
}}_{\nu }\ ,
\end{equation}
\begin{equation}
{\eta }^{+}\equiv \prod\limits_{\mu  =0}^{ 3} {\tilde{\Delta }}_{\mu }\  \ ,\ \ {\eta }^{-}\equiv \prod\limits_{\mu 
=0}^{ 3} {\tilde{\nabla }}_{\mu }\ .
\end{equation}

From the Fourier transform we can derive the wave equation in lattice space.

\noindent {\it Type I Fourier transform:}

The kernel of the transform satisfies:
\begin{equation}
{\frac{1}{\varepsilon }}{\Delta }_{\mu }\exp\ \left(2\pi  i\left({k.j}\right)\varepsilon\right)  =i{\frac{2}{\varepsilon }}\tan\ \left(\pi
{ k}_{\mu }\varepsilon\right) {\tilde{\Delta }}_{\mu } \exp\ \left(2\pi  i\left({k.j}\right)\varepsilon\right) 
\end{equation}

We could apply the Fourier transform to the Dirac equation in momentum space (59) and would obtain the discrete wave equation.
 Instead we postulate a difference equation that in the limit goes to the continuous differential equation, namely,
\begin{equation}
\left({i{\gamma }^{\mu }{\delta }_{\mu }^{+}-{m}_{0}c{\eta }^{+}}\right)\psi \left({{ j}_{\mu }}\right)=0.
\end{equation}

The kernel of the Fourier transform (25) or ``plane wave'' is a particular solution of (67) if it satisfies
\begin{equation}
\left({{\gamma }^{\mu }{\frac{2}{\varepsilon }}\tan\pi { k}_{\mu }\varepsilon  -{m}_{0}c}\right){\eta }^{+}\ \exp\ 2\pi
 i\left({k\cdot  j}\right)\varepsilon  =0.
\end{equation}

Applying the operator ${\gamma }^{\mu }{\frac{2}{\varepsilon }}\tan\pi { k}_{\mu }\varepsilon  +{m}_{0}c$ from the left
to the last equation we obtain
\begin{equation}
{\frac{4}{{\varepsilon }^{2}}}\left({\tan\ \pi { k}^{\mu }\varepsilon }\right)\left({\tan\ \pi { k}_{\mu }\varepsilon
}\right)-{m}_{0}^{2}{c}^{2}=0,
\end{equation}
 which is the integrability condition for the solution of the wave equation.

Now we multiply eq. (68) by some arbitrary (periodic) function $\psi \left({{ j}_{\mu }}\right)$ of discrete variable and
sum for all $j_{\mu}.$

We get
\begin{equation}
\left({{\gamma }^{\mu }{\frac{2}{\varepsilon }}\tan\pi { k}_{\mu }\varepsilon  -{m}_{0}c}\right)\hat{\psi
}\left({{k}_{\mu }}\right)=0,
\end{equation}
where
\begin{equation}
\hat{\psi }\left({{k}_{\mu }}\right)=\sum\limits_{{j }_{\mu }=0}^{ N-1} \psi \left({{ j}_{\mu }}\right){\eta }^{
+} \exp\ \left(i2\pi { k}^{\mu }{ j}_{\mu }\varepsilon \right) 
\end{equation}
is the Fourier transform of $\psi \left(j_{\mu }\right)$.

If we compare eq. (70) with eq. (59) both are identical if we restrict the momentum $p_{\mu}$ to the discrete values
\begin{equation}
{p}_{\mu }={\frac{2}{\varepsilon }}\tan\pi { k}_{\mu }\varepsilon  \ \ ,\ \ {k}_{\mu }={\frac{1}{2\pi  i\varepsilon
}}\ln{\frac{1+{\frac{1}{2}}i\varepsilon { p}_{\mu }}{ 1-{\frac{1}{2}}i\varepsilon { p}_{\mu }}}
\end{equation}

Instead of postulating Dirac equation on the lattice (67) we can deduce it in a natural way from the projection operator of the
Dirac representation (59): first identify the momentum $p_{\mu}$ in this expression with the new variable $k_{\mu}$ as given in
(72), then apply the Fourier transform of type I (23) and finally, using (66), obtain (67).

Applying to the wave equation the operator $i{\gamma }^{\mu }{\delta }_{\mu }^{-}+{m}_{0}c{\eta }^{-}$ from the left we obtain
the discrete version of the Klein-Gordon equation in the lattice space
\begin{equation}
\left({{\delta }_{\mu }^{+}{\delta }^{\mu \rm -}-{m}_{0}^{2}{c}^{2}{\eta }^{+}{\eta }^{-}}\right)\psi \left({{j}_{\mu
}}\right)\rm =0
\end{equation}
a particular solution of which is again the ``plane wave'' (25) provided the integrability conditions (69) is satisfied

\noindent {\it Type II Fourier transform:}

The kernel of this Fourier transform satisfies:
\begin{equation}
{\frac{1}{\varepsilon }}{\Delta }_{\mu }\ \exp{\frac{2\pi i}{N}}\left({j \cdot m}\right)=i{\frac{2}{\varepsilon }}\tan\
\left({\frac{\pi }{N}}m\right){\tilde{\Delta }}_{\mu }\ \exp{\frac{2\pi i}{N}}\left({j \cdot m}\right)
\end{equation}

We postulate a difference equation on the lattice space the limit of which goes to the continuous wave equation,
\begin{equation}
\left({i{\gamma }^{\mu }{\delta }_{\mu }^{+}-{m}_{0}c{\eta }^{+}}\right)\psi \left({{j}_{\mu }}\right)\rm =0
\end{equation}

The kernel of the Type II Fourier transforms (38) is a particular solution or ``plane wave'' of this equation (75) if it
satisfies:
\begin{equation}
\left({{\gamma }^{\mu }{\frac{2}{\varepsilon }}\tan\ {\frac{\pi }{N}}{m}_{\mu }-{m}_{0}c}\right){\eta }^{+}\exp{\frac{2\pi i
}{N}}\left({j\cdot m}\right)=0
\end{equation}

Applying the operator ${\gamma }^{\mu } \frac{2}{\varepsilon }\tan\ {\frac{\pi }{N}}{m}_{\mu }+{m}_{0}c$ from the left we
get
\begin{equation}
{\frac{4}{{\varepsilon }^{2}}}\left({\tan\ {\frac{\pi }{N}}{m}_{\mu }}\right)\left({\tan\ {\frac{\pi }{N}}{m}_{\mu
}}\right)-{m}_{0}^{2}{c}^{2}=0,
\end{equation}
which is the integrability condition for the ``plane wave solution'' of the wave equation on the lattice.

Now we multiply eq. (71) by some arbitrary (periodic) function $\psi \left({j_{\mu }}\right)$ of discrete variable and
sum for all $j_{\mu }$. We get
\begin{equation}
\left({{\gamma }^{\mu }{\frac{2}{\varepsilon }}\tan\ {\frac{\pi }N}{m}_{\mu }-{m}_{0}c}\right)\hat{\psi }\left({{k}_{\mu
}}\right)=0,
\end{equation}
where 
\begin{equation}
\hat{\psi }\left({{k}_{\mu }}\right)=\sum\limits_{j_{\mu }=0}^N \psi \left({j_{\mu
}}\right){\eta }^{\rm +}{\rm exp}\ i{\frac{2\pi }N}{m}^{\mu }{j}_{\mu }
\end{equation}
is the Fourier transform of $\psi \left({j_{\mu }}\right)$.

Both eq. (59) and eq. (78) are identical if we restrict the momentum $p_{\mu}$ to the discrete values
\begin{equation}
{p}_{\mu }={\frac{2}{\varepsilon }}\tan\ {\frac{\pi }{N}}{m}_{\mu }\ ,\ {m}_{\mu }={\frac{N}{2\pi 
i}}\ln{\frac{1+{\frac{1}{2}}i\varepsilon {p}_{\mu }}{1-{\frac{1}{2}}i\varepsilon {p}_{\mu }}}\ \ \ {m}_{\mu
}=0,1,\cdots \rm ,N-1
\end{equation}

From eq. (75) we could derive the Klein-Gordon equation on the lattice space as in eq. (73) with the integrability conditions
(69) as before.

As in the case of Fourier transform of type I we can deduce again the wave equation on the lattice in a natural manner from the projection
operator (59): identify there the momentum ${p}_{\mu }$ with the new variable given by (80), use the Fourier transform (37) of type II,
together with (74) and obtain (75).

\section{Induced representations of the discrete Poincar\'e groups}

Let ${ \cal P}_{ +}^{\uparrow } ={T}_{4}{\times}_{s}SO\left({3,1}\right)$ be the Poincar\'e group
restricted to the integral Lorentz transformations and discrete traslations on the lattice with
the group composition
\begin{equation}
\left({ a,\wedge }\right)\left({ a',\wedge '}\right) =\left({a+\wedge a',\wedge \wedge
'}\right).
\end{equation}
In order to construct irreducible representations we follow the standard method [13]: 

(1) Choose an $UIR,\ {D}^{\stackrel{o}{k}}\left({a}\right),$ of the translation group $T_4$

(2) Define the little group $H \in  SO(3,1)$ by the stability condition
\begin{equation}
h\in H\ \ :\ \ {D}^{\stackrel{o}{k}}\left({{h}^{-1}a}\right)={D}^{\stackrel{o}{k}}\left({a}\right)
\end{equation}

(3) Choose an $UIR\ D^{\alpha}$ of the little group $H$ and construct for the group $T_4\times_sH$ the $UIR$ 
\begin{equation}
{D}^{\stackrel{o}{k} ,\alpha }\left({a,h}\right)
={D}^{\stackrel{o}{k}}\left({a}\right)\otimes {D}^{\alpha }\left({h}\right)
\end{equation}
(4) Choose coset generators $c$ of $T_4\times_sH$ in ${ \cal P}_{ +}^{\uparrow }$ constructed from the group action
\begin{equation}
\left({\tilde{a},\tilde{\wedge }}\right)c=c'\left({a,h}\right)
\end{equation}
(5) Then the induced representations is: 
\begin{equation}
{D}^{\stackrel{o}{k},\alpha }_{c'c}\left({\tilde{a},\tilde{\wedge
}}\right)={D}^{\stackrel{o}{k},\alpha }\left({a,h}\right)\delta
\left({{\left({c'}\right)}^{-1}\left({\tilde{a},\tilde{\wedge
}}\right)c,\left({a,h}\right)}\right)
\end{equation}
and this is an $UIR$ of ${\cal P}_{+}^{\uparrow }.$

Recall the construction of the massive $UIR$ for the continuous case: If $\stackrel{o}{k}$ is any vector inside the forward light cone, one
can shift it on the orbit by a continuous Lorentz transformation to the form $\stackrel{o}{k}=m_0c \left({1,0,0,0}\right)$, with the stability
group $H=SO(3)$. In the $2\times2$ matrix form this stability group becomes $SO(2)$.

For the discrete Poincar\'e group, we may choose $\stackrel{o}{k}$ from the intersection of the Brillouin zone with the forward light cone. We
can shift it on the orbit only by discrete Lorentz transformations. Certainly we can choose  $\stackrel{o}{k}=m_0c \left({1,0,0,0}\right)$
within the Brillouin zone. Then the discrete stability group is the cubic group, and in the $2\times2$ matrix form it becomes the binary cubic
group. In both cases the representations subduced from the continuous to the discrete little groups remain irreducible.

For the coset representative $c\equiv \left({0,\wedge }\right)$ we can choose the integral
Lorentz transformations $\wedge \equiv L\left({k}\right)$ that take $\stackrel{o}{k}$
into an arbitrary integral vector of the unit hyperboloid. These transformations were defined in section 2 as integral boosts.
The Dirac delta function in (85) is zero unless
\[\left({a,h}\right)
=\left({0,{L}^{-1}\left({k'}\right)}\right)\left({\tilde{a},\tilde{\wedge
}}\right)\left({0,L\left({k}\right)}\right)=\left({{L}^{-1}\left({k'}\right)\tilde{a},{L}^{-1}\left({k'}\right)\tilde{\wedge
}L\left({k}\right)}\right)\]
Substituting in eq. (85) with $c,c'\rightarrow  k,k'$ and using (82) we get 
\begin{equation}
{D}_{k'k}^{\stackrel{o}{k},\alpha }\left({\tilde{a},\tilde{\wedge }}\right)
={D}^{\stackrel{o}{k}}\left({{L}^{-1}\left({k'}\right)\tilde{a}}\right){D}^{\alpha
}\left({{L}^{-1}\left({k'}\right)\tilde{\wedge }L\left({k}\right)}\right)
\end{equation}
The spinor representation of the second factor is given with respect to the element
${L}^{-1}\left({k'}\right)\tilde{\wedge }L\left({k}\right)$ that belongs to the little group,
$SU(2)$, or $SO(3,R)$ respectively. These representations of $SU(2)$ corresponding to spin 1/2 or spin 1 state irreducible when
restricted to the (binary) cubic group [11]. The first factor can be written: 
\begin{equation}
{D}^{\stackrel{o}{k}}\left({{L}^{-1}\left({k'}\right)\tilde{a}}\right)
={D}^{k'}\left({\tilde{a}}\right)
\end{equation}
where $k'={\left({{L}^{-1}\left({k'}\right)}\right)}^{T}\stackrel{o}{k}$ are all the points that defined
the $UIR$ of the translation group and belong to the orbit on the dual translation group. This orbit is discrete in our
analysis.

We apply the analysis given in [13] for the semidirect product of the discrete translation group and a point group on the
lattice. The dual group of the translation group is given by all the points from the Brillouin zone. We wish to characterize
the discrete orbit of the point group by function on the dual space. We formulate five conditions for these contraints:

(1) they should vanish on the orbit points,
(2) they should admit a periodic extension on the $k$-space,
(3) the constraints must be Lorentz invariant,
(4) the constraints should vanish only on the points of the orbit,
(5) when the lattice spacing goes to zero, the difference equations in position space should go the continuous wave equation
in Minkowski space.

If we require only condition (1)-(4) the polynomials $\left({{k}^{\mu }{k}_{\mu }-{\stackrel{o}{k^{2}}}}\right)=0$ vanish on and
only on the discrete points of the orbit and the irreducible representation could be characterized by the functions (see Ref.
[15], p. 192)
\begin{equation}
f{}^*\left({k}\right):\left({{k}^{\mu }{k}_{\mu }-{\stackrel{o}{k^{2}}}}\right)f{}^*\left({k}\right)=0.
\end{equation}
Nevertheless the difference equation (73) we have constructed in section 5, with the kernel of Fourier transform of Type I and
II, does not lead to this constraint. We take a new approach. We choose the constraints we have derived in the Dirac
representation in momentum space eq. (60) for the continuous case:
\begin{equation}
\left({{p}^{\mu }{p}_{\mu }-m_{0}^2c^2}\right)\psi \left({p}\right)=0
\end{equation}

If we identify ${p}_{\mu }={\frac{2}{\varepsilon }}\tan\ \pi {k}_{\mu }\varepsilon $, with $k$ in the Brillomin zone and
$\varepsilon$ the lattice spacing and use the Fourier transform of Type I we obtain the difference equation (73)

If we identify ${p}_{\mu }={\frac{2}{\varepsilon }}\tan\ {\frac{\pi }{N}}{m}_{\mu }$ in (89) and use the Fourier transform of
Type II we obtain again the difference equation (73), the continuous limit of which leads to the continuous Klein-Gordon
equation.

The constraints (69) are periodic with respect to the $k_{\mu}$-space in the Type I and the constraints (77) are periodic with respect the
$m_{\mu}$-space in the Type II.

Nevertheless when integral Lorentz transformations are applied to the components of the $k_{\mu}$ or $m_{\mu}$ variables the
new $p_{\mu}$ do not satisfy the constraint equations, and at the same time the constraints vanish at points not on the orbit.
Therefore conditions (3) and (4) are violated, although they can be recovered in the asymptotic limit when $\varepsilon
\rightarrow 0$
\begin{equation}
\left({{k}^{\mu }}{{k}_{\mu }}-{\stackrel{o}{k^{2}}}\right)\psi \left({{k}_{\mu }}\right)\rm =0\ \ {\rm Type\
I}
\end{equation}
or in the limit $\varepsilon \rightarrow 0$ \ ,\ \ $N\varepsilon \rightarrow 2 \pi$.
\begin{equation}
\left({{m}^{\mu }{m}_{\mu }-1}\right)\psi \left({{m}_{\mu }}\right)=0\ \ {\rm Type\ II}.
\end{equation}
The corresponding representation becomes irreducible and invariant.

In both cases the wave equation in lattice space gives in the assymtotic limit the continuous Dirac equation.

\vskip 1truecm  \noindent  { \bf Acknowledgments}
\vskip 0.5cm

One of the authors (M.L.) wants to expressed his gratitude to Profesor Reinhardt, the Director of the Institut
f\"{u}r  theoretische Physik, Universit\"{a}t Tubingen, where part of this
work was done, for the hospitality. This work has been partially supported by D.G.I.C.Y.T. contract
\#Pb94-1438 (Spain).


\begin{thebibliography}{00}
\bibitem{} A. Schild, ``Discrete space-time and integral Lorentz transformation'', Can. J. Math.
{\bf 1}, 29 (1948).
\bibitem{} V. Kac, {\it Infinite dimensional Lie Algebras}, Cambridge U. Press (1991)pp. 69-71.
\bibitem{} M. Lorente, ``Cayley parametrization of semisimple Lie groups and its application to
Physical Laws in a (3+1)-dimensional cubic lattice'', Int. J. Theor. Phys. {\bf 11}, 213-247
(1974). ``A realistic interpretation of lattice gauge theories'' {\it Fundamental Problems in Quantum Physics} (M. Ferrero, A.
van der Merwe, ed.) Kluwer 1995, p. 177-186.
\bibitem{} C. M{\o}ller, {\it The theory of Relativity}, Oxford Clarendon Press, 1952, p. 42.
\bibitem{} Ref. 3, p. 221.
\bibitem{} M. Lorente, ``A new scheme for the Klein-Gordon and Dirac fields on the lattice with
Axial Anomaly'', {\it J. Group Th. in Phys.}. {\bf 1} 105-121 (1993), p. 107.
\bibitem{} M. Creutz, {\it Quarks, gluons and lattices}, Cambridge U. Press, 1983, p. 15. See
also, I. Montvay, G. M\"{u}nster {\it Quantum Field on the lattice}, Cambridge U. Press, 1994.
\bibitem{} M. Lorente, ``Discrete Reflection Groups and Induced Representations of Poincar\'e
Group on the Lattice'' {\it Symmetries in Science IX} (B. Gruber, M. Ramek ed.) Plenum, N.Y. 1997, p. 211-223.
\bibitem{} L. Fonda, G.C.Ghirardi, {\it Symmetry Principles in Quantum Physics}, Marcel Dekker
1970, p. 309.
\bibitem{} U.H. Niederer, L.O'Raifertaigh, ``Realization of the Unitary Representations of the
Inhomogeneous Space-time groups'', I and II, Forsch. Phys {\bf 22}, 111-129, 131-157 (1974).
\bibitem{} Melvin Lax, ``Symmetry principles in Solid State and Molecular Physics'', John Wiley \& sons, New York
1974, p. 431-2, 436-8.
\bibitem{} I. Montvay, ``Supersymetric gauge theories on the lattice'', Lattice 96, Nuclear
Physics B (Proc. Suppl.) {\bf 53} (1997), 853-5.
\bibitem{} P. Kramer, M. Lorente, ``Discrete and continuous symmetry via, induction and duality'',
{\it Symmetries in Science X}. (B. Gruber, M. Ramek ed.) Plenum, N.Y. 1998, p. 165-177.
\bibitem{} P. Slodowy, ``Simple Singularities and Simple Algebraic Groups'', Lecture Notes in Math. {\bf 815},
Springer (1980), p. 70-75.
\bibitem{} M. Lorente, P. Kramer, ``Induced Representations of the Poincar\'e group on the lattice: spin 1/2 and 1
case'', {\it Symmetries in Science X} (B. Gruber, M. Ramek ed.) Plenum, N.Y., p. 179-195.
\end{thebibliography}
\end{document}